# Discussion of crowd evacuation models organized according to symmetry analysis approach

W. Sikora, J. Malinowski

Faculty of Physics and Applied Computer Science, University of Science and Technology, al.Mickiewicza 30, 30-059 Krakow, Poland

**Abstract** The evacuation of football stadium scenarios are discussed as model realizing ordered states, described as movements of individuals according to fields of displacements, calculated correspondingly to given scenario. The symmetry of the evacuation space is taken into account in calculation of displacements field - the displacements related to every point of this space are presented in the coordinate frame in the best way adapted to given symmetry space group, which the set of basic vectors of irreducible representation of given group is. The speeds of individuals at every point in the presented model have the same quantity. As the results the times of evacuation and average forces acting on individuals during the evacuation are given. Both parameters are compared with the same ones got without symmetry considerations. They are calculated in the simulation procedure. For realization the simulation tasks the new program (using modified Helbing model) was elaborated.

**Introduction**

The behavior of social systems under the action of some external conditions may be considered in analogy to the behavior of solid states under the action of temperature or external electric or magnetic fields. Both systems are complex, containing many interacting elements, and are realized in strictly defined spaces. Very often these spaces are strongly restricted, and because of these restrictions not any types of evolutions of these systems are allowed. When these spaces are symmetrical (crystals are good examples of such situation) the symmetry considerations conducting in the frame of theory of groups and representations are able to predict the types of behavior of the systems, which are permitted by the symmetry of these spaces. The symmetry analysis method (SAM) [1] from many years successfully was used for significant simplification of descriptions of different type of phase transitions in crystals. The first trial of application the SAM to modeling the evacuation of one floor of skyscraper, in which the symmetry of allowed space had been broken because of some accident, is presented in our work at KES AMSTA [2] It had been done, that evacuation along the paths related to the field of displacements calculated by SAM in agreement with symmetry of allowed space leads to a little longer time of evacuation (because they are not the shortest ways to the exits) but significantly lower average force acting on evacuated individuals. . It is known**,** that to big press acting on evacuated agents during evacuation occurs very often the cause of tragic accidents, so the possibility to lower this press seems to be important. .Taken into account the crowd psychology is rather not realistic assumption, that in empty space of the floor evacuated individuals choose the paths to the allowed exits non along shortest way, but along the line indicated by SAM, even they would be drown in some maner by some AMI devices. There are such constructions, where many people are keeping together, where exists the possibility to build the paths of evacuation in agreement with SAM field of displacements. Such construction is for example football stadium. In this work the model of evacuation of one floor of skyscraper regarded in [2] is applied to one sector of football stadium. The football stadium build from many similar sectors is a very good approximation of the structure fulfilling the periodic boundary conditions. The results of symmetry calculations done for one floor of skyscraper may be applied also to planning the evacuation paths on stadium sector. We discuss the evacuation of stadium which architecture takes into account the symmetry calculations and compare it with evacuation of traditional stadium.

**Elements of the symmetry analysis method**

The symmetry group of some object is defined as some special set of transformations which leave this object unchanged. As the elements of space groups appear translations, mirror planes, inversion and some special rotations around the axis, as good as all combinations of them. Translational symmetry gives the possibility to choose the small block named elementary cell. The repetition of it reconstructs all system. In this way description of all system is reduced to description of elementary cell. Such procedure reduces the number of parameters needed to description of the system, but steel this number may be decreased by using full symmetry restrictions, not only translational one. The properties of given system in mathematical language are presented as functions defined on allowed space. In crystals it may be the displacements of atoms from initial equilibrium positions, magnetic moments localized on given positions or probability of sites occupations for example. The transformation properties of such functions are given by matrix representations of allowed space symmetry group. One may find the books where the representations of different groups (also crystallographic 32 point groups and 230 space groups) are calculated and listed.

Theory of group representations had been applied for simplification of description of many-body, complex physical systems many years ago. E. Wigner, G.J.Lubarski and A.P. Cracknell introduced as "symmetric coordinates" set of basis vectors of irreducible representations of molecules symmetry groups in calculations of molecules vibrations. Lubarski discussed also the role of irreducible representations (IR) of crystal symmetry group in crystallographic second order phase transitions. This method to the description of magnetic ordering in crystals was at first introduced by E.F. Bertaut. Later that line of analysis has been developed by many other theoreticians, like Izyumov and others [1]. The presentation of the functions describing interesting system properties, localized on given set of symmetry equivalent positions, in the usually used frame of coordinates related to the edges of elementary cell (crystallographic system), as was mentioned previously, takes advantage of translation symmetry only. The other symmetry relations are lost in this description and as a consequence the description of many system properties is not as simple as possible. From rules of the theory of group representations follows that any function defined on the space with given symmetry group may be presented as linear combination of basis vectors of irreducible representations (BV-s) $\Psi$ of this group:

$$\mathbf{S} = \sum_{l,\nu,\lambda} c_\lambda^{\mathbf{k}_l,\nu} \Psi_\lambda^{\mathbf{k}_l,\nu} \quad (1)$$

(l - number of **k** vectors describing the translational symmetry of ~S, ν- number of IR's, λ - number of dimensions of ν's IR).

From equation (1) follows, that the function S may be treated as some vector in the functional space, the BV-s $\Psi$ play the same role in the presentation of function S, as axis of coordinate system, and coefficients $c^{k_l,\nu}_\lambda$, as components of it. When irreducible representation is taken into account the dimension of this functional space invariant under all group transformations is the smallest one. It leads to smallest number of $c^{k_l,\nu}_\lambda$ parameters, which in the frame of BV-s $\Psi$ presents the function S. Each choice of these free parameters $c^{k_l,\nu}_\lambda$ uniquely determines one of the possible models of new structure that may be realized after the phase transition. The presentation of model structures in the frame of basis vectors of irreducible representations of the initial symmetry group instead of that in the frame of crystallographic system (x; y; z), is the best matching to the symmetry of the problem and it provides the simplest (requiring the lowest number of independent parameters) form of the structure description.

The form of the basis vectors and the information which of the representations take part in the phase transition under consideration are directly given by the theory of groups and representations. It is important to note that the basis vectors have the same translational properties as Bloch functions. Therefore, the basis vectors may be defined on positions in given elementary cell of the crystal as well as in the elementary cell translated by a lattice vector **t**, which just corresponds to a multiplication by $e^{ikt}$. Not all from the possible $c^{k_L v_L}$ are allowed, because the parameters should be selected in such way that the resulting properties related to all atoms have real values. This condition influences on the set of equations which $c^{k_L v_L}$ have to satisfy and as a result the number of independent free parameters is reduced and strictly determined. After such operation the final model contains clearly defined minimum number of free parameters and presents strictly defined relations between localized on different crystal sites quantities describing considered property. The choice of representation $\tau_v$ and the coefficients $c^{k_L v_L}$ uniquely determines the symmetry of the new, ordered structure, independently on the kind of the property taken into account. The type of phase transition and the property under consideration is included in the form of basis vectors. In the frame of SAM also discussion of connections between different types of transformations is possible.

For realization of the mentioned problems the computer program MODY based on the representations of symmetry space groups [2], and another associated programs - SPLIT, Tensor-vis, Tensor_OpenGL, had been elaborated. They offer to find all possibilities of new scalar, vector and tensor type structure properties allowed by the initial symmetry of the structure.

The discussion of coexistence of different types of system behavior, leading to different properties of the system are based on the assumption, that different functions describing these properties of the system should have the same symmetry. From theory of representation it is known, that these different functions should belong to the same irreducible representation of system symmetry group.

**Application the SAM to stadium evacuation scenario**

The scenario of evacuation of football stadium organized according to SAM approach is discussed in this chapter and compared with traditional model of stadium. By using the simulations with and without implementation of SAM results the time of evacuations and average forces acting on the pedestrians during the evacuation are calculated.

The football stadium, similar as existing in Munich for example, are the space where the big number of individuals are localized, which with quite good approximation presents the translational symmetry with periodic boundary conditions (usually taken into account in crystals). This is the reason that as the first step of stadium evacuation scenario the evacuation of one stadium sector may be discussed. It significantly reduces the number of evacuated individuals and, as consequence, the number of parameters used for scenario description. The real symmetry of the sector is described by P1m1 space group. For designation the ordering of chairs and evacuation paths guaranteeing quick and safe evacuation, the field of displacements which agree with full symmetry of available space should be calculated by using the vector type SAM. On the base of P1m1 symmetry group the number of positions in the sector, which should be regarded is reduced to ½, but the directions of displacements which should be defined on these positions remain free. The best choice of displacement directions from evacuation point of view can be made by regarding the P1m1 group as the broken symmetry of Pmm2 group generated by τ₄ representation of **k**=(0,0,0), which lead to field of displacements on the xy plane (which are the polar vector quantities), and corresponding field of speeds (which are the scalar quantities) as given in the Table1 quoted below. Because the steps going up to exits, the displacemenst at each positions should have the z components, directed to the up. As it is shown in the Tab1, such relation between z components belongs to the $\tau_1$ representation. From SAM follows that $\tau_1$ may be associated with any representation, because it doesn't change the initial symmetry. Taking into account the vector field of displacements got by mentioned above treatment the paths for evacuation are design in one stadium sector.

Table1

| Representation and destination group | Version | Free parameter | The polar vector type and scalar type basic vector functions of representations of Pmm2 space group with k=(0,0,0), defined at positions 4i in the elementary cell | | | |
|---|---|---|---|---|---|---|
| | | | Pos. 4i | | | |
| | | | 1: (x, y, 0) | 2: (1-x, 1-y,0) | 3: (1-x, y, 0) | 4: (x, 1-y,0) |
| Tau1 destination group Pmm2 | vectorI | A₁ | (1,0,0) | (-1,0,0) | (-1,0,0) | (1,0,0) |
| | vectorII | B₁ | (0,1,0) | (0,-1,0) | (0,1,0) | (0,-1,0) |
| | vectorIII | C₁ | (0,0,1) | (0,0,1) | (0,0,1) | (0,0,1) |
| | scalar | P1 | 1 | 1 | 1 | 1 |
| Tau2 destination group P112 | vectorI | A₂ | (1,0,0) | (-1,0,0) | (1,0,0) | (-1,0,0) |
| | vectorII | B₂ | (0,1,0) | (0,-1,0) | (0,-1,0) | (0,1,0) |
| | vectorIII | C₂ | (0,0,1) | (0,0,1) | (0,0,-1) | (0,0,-1) |
| | scalar | P2 | 1 | 1 | -1 | -1 |
| Tau3 destination group Pm11 | vectorI | A₃ | (1,0,0) | (1,0,0) | (-1,0,0) | (-1,0,0) |
| | vectorII | B₃ | (0,1,0) | (0,1,0) | (0,1,0) | (0,1,0) |
| | vectorIII | C₃ | (0,0,1) | (0,0,-1) | (0,0,1) | (0,0,-1) |
| | scalar | P3 | 1 | -1 | 1 | -1 |
| Tau4 destination group P1m1 | vectorI | A₄ | (1,0,0) | (1,0,0) | (1,0,0) | (1,0,0) |
| | vectorII | B₄ | (0,1,0) | (0,1,0) | (0,-1,0) | (0,-1,0) |
| | vectorIII | C₄ | (0,0,1) | (0,0,-1) | (0,0,-1) | (0,0,1) |
| | scalar | P4 | 1 | -1 | -1 | 1 |

**Simulation of stadium evacuation**

The evacuation of the stadium sector is simulated by using the Helbing "social forces" model [4], which was taken as a starting point to model presented in this paper.

$$m_i \frac{d\vec{v}_i}{dt} = m_i \frac{v_i^0 \vec{e}_i^0(t) - \vec{v}_i(t)}{\tau} + \sum_{j(\neq i)} \vec{f}_{ij} + \sum_W \vec{f}_{iW} \quad (2)$$

Equation (2) on right side has a sum of three forces acting on human during evacuation simulation. Second and third one describe sum of repel forces from all other mans and sum of repel forces from all walls and all obstacles inside the evacuated space. First force is the most interesting at this stage. In general it describes in which direction each person should go. In simple case like empty room with some doors, this force can be easily calculated as a sum of attract forces from all doors. Here is also the place for the implementation of vector field calculated by symmetry considerations. This assumption is inadequate with more complicated geometry of the building. Figure (1) shows the example when such approach leads to situation when pedestrian will stick for eternity in dead end.

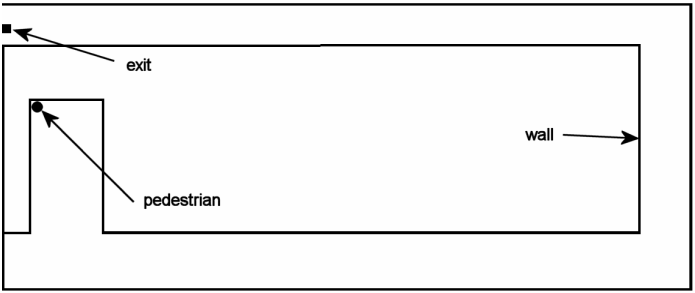

Fig 1. The „dead-end".

In approach presented in this paper floor space was divided into cells (this got nothing in common with "cellular automata" approach to such kind of simulations!). Every cell can contain obstacle, free space or exit. Every cell that contain free space have vector of desirable velocity connected with it, calculated according to schema described below. Any person that stands inside particular cell takes this vector as his desirable velocity. Unlike in cellular automata approach in this approach "cell" can contain one or more persons and any direction of velocity. Single cell size depends on building geometry only. To make a simulation of an evacuation of single room, small number (like 7x6 for example) of cells is more than enough. Figure (2) show example of such table presenting the final field of velocities. These vectors can be calculated using "ray casting" method as shown in paper [5].

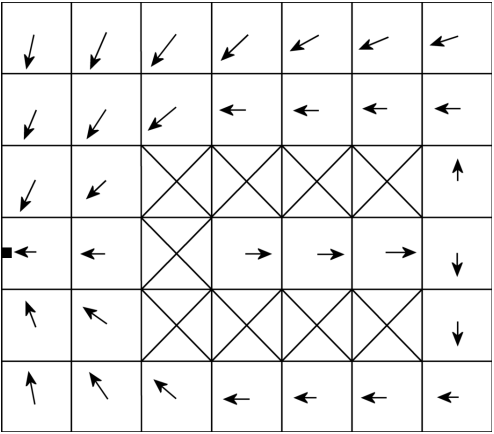

Fig 2. Final field of velocities

The vector field calculated for stadium sector by SAM is presented on the figure 3 and the sector architecture which takes into account these SAM calculations for building the evacuation paths is presented on the figure 4. It gives the map of free spaces, obstacles and exits needed for calculation of forces acting on individuals as mentioned in the equation 2 and velocities field used in the simulation (figure 5). The traditional stadium sector architecture is given on the figure 6.

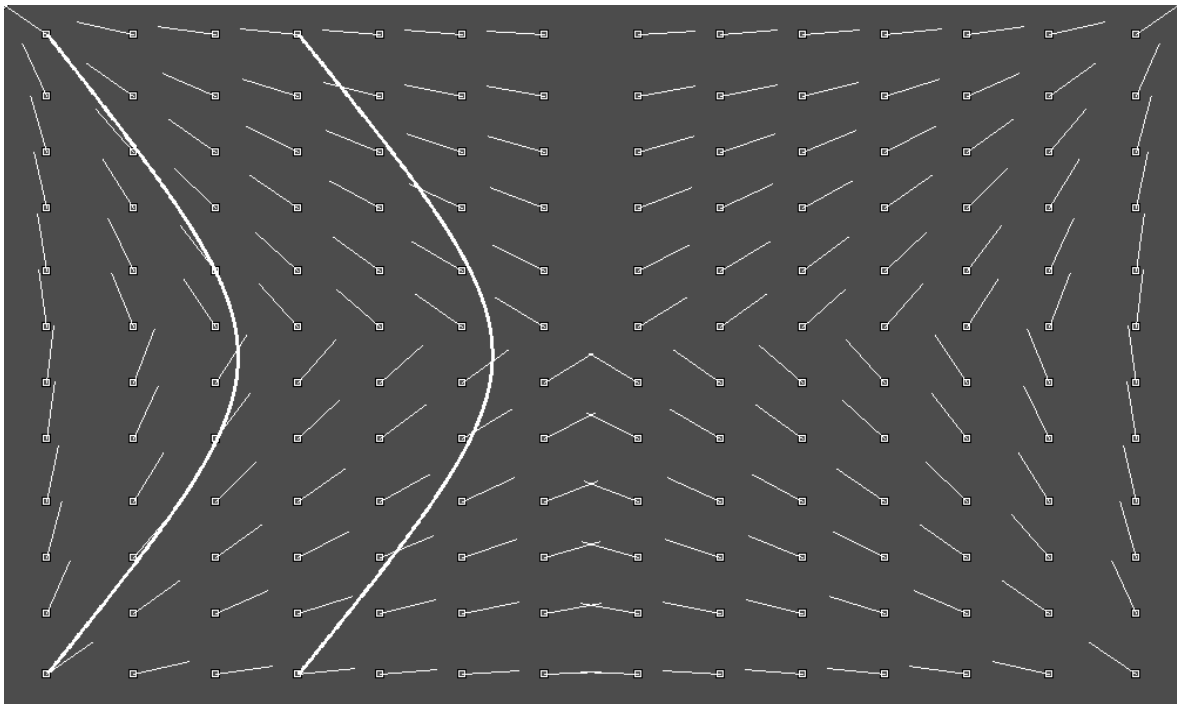

Figure 3 The vector field calculated by SAM related to one stadium sector for Pmm2 space group and $\tau_4$ representation of **k**=(0,0,0)

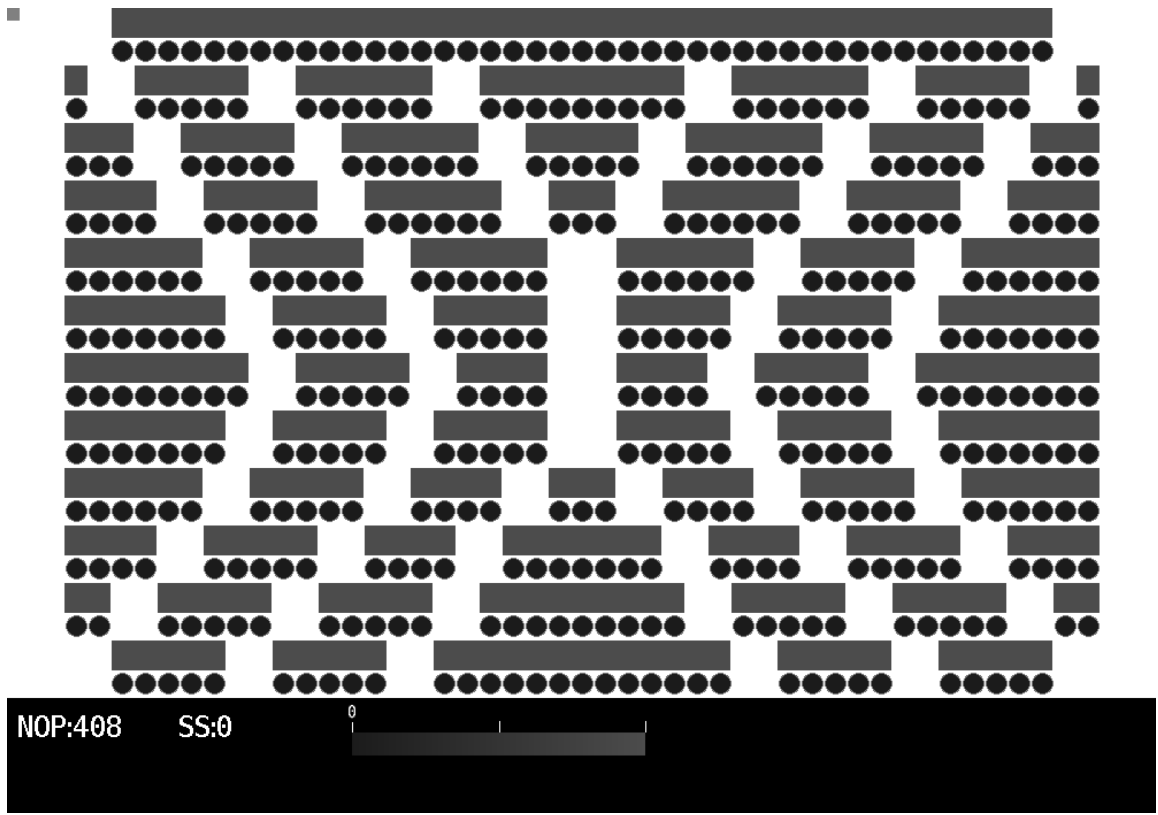

Figure 4 Construction of evacuation paths in stadium sector with agreement to SAM vector field calculations

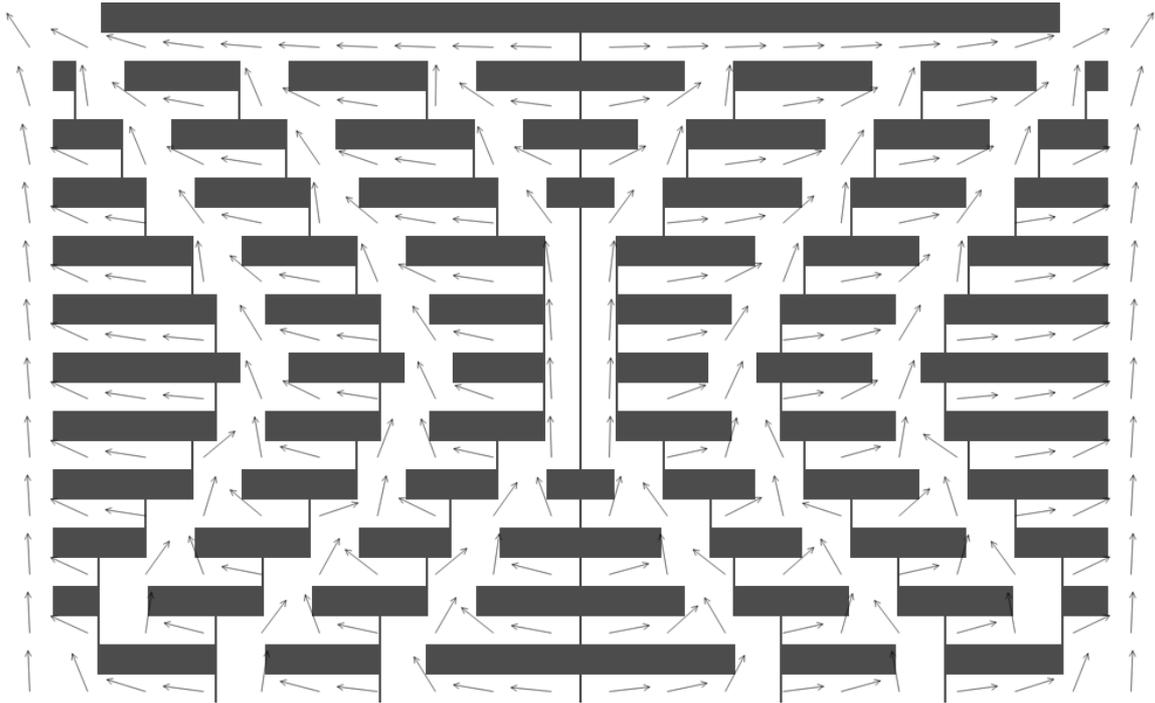

Figure 5   Table of final field of velocities calculated for simulation of stadium sector evacuation for SAM model

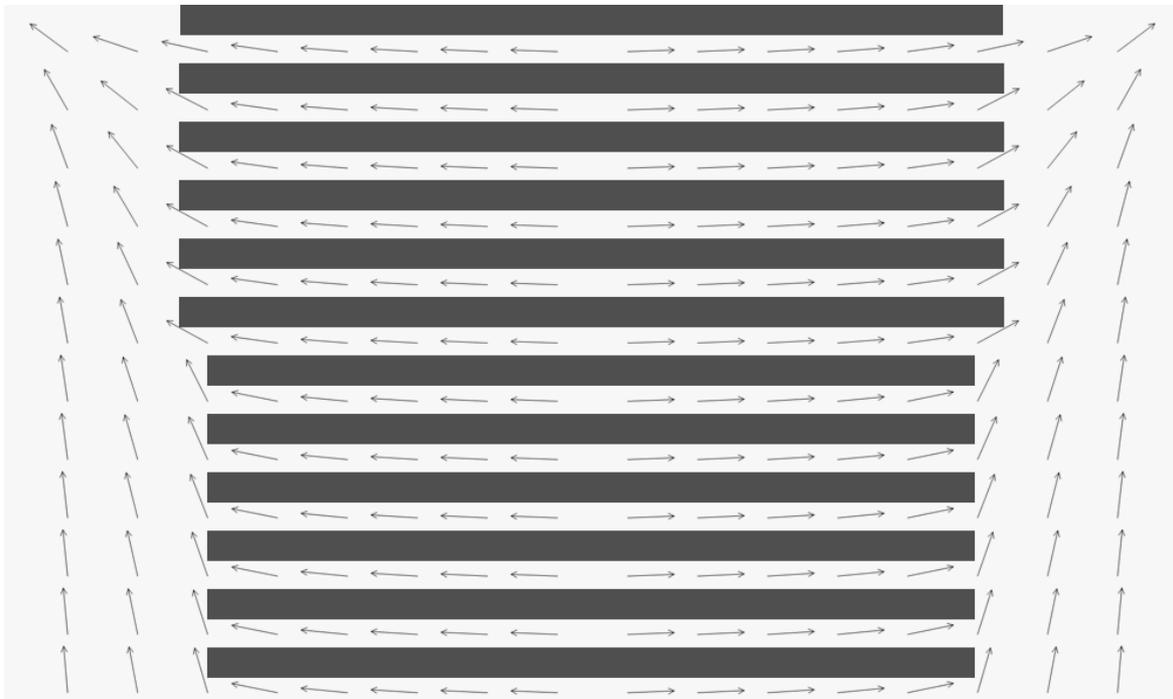

Figure 6  Table of final field of velocities calculated for simulation of stadium sector evacuation for traditional model

The time of evacuation and average forces acting on the individuals are calculated during the simulation by using SAM related and traditional architectures of stadium. The number of evacuated individuals – 408 - taken for simulation agrees with number of sites in the sector of Munchen football stadium and is the same for both analysed architectures. This two models differ only in distribution of chairs and evacuation paths inside the sector. The results for both models – SAM related and traditional - are given as the plots presenting the average force acting on one individual as the function of time from the beginning to the end of evacuation. The comparison of these functions is seen on the figure 7.

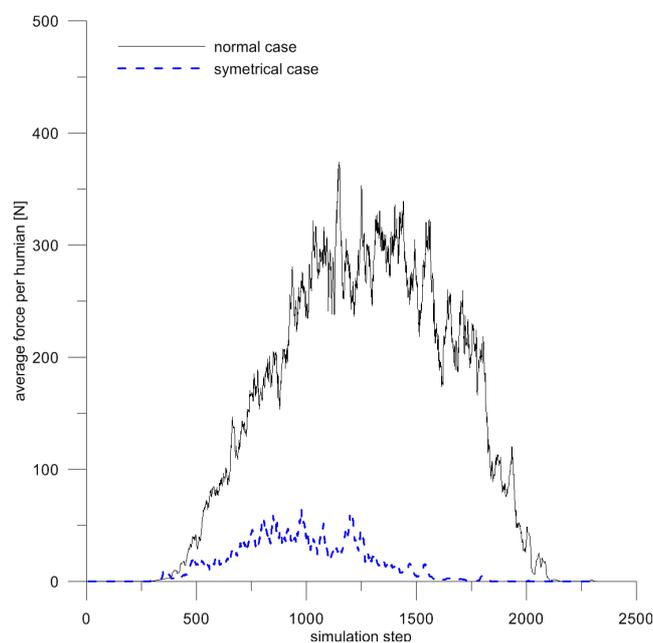

Figure 7. Average forces acting on pedestrian at each step of evacuation process calculated for traditional and SAM stadium models.

Conclusions

The new model of architecture of sectors of football stadium is proposed, which leads to the best sector evacuation. As the criterion of goodness of evacuation two parameters - the time of evacuation $t_e$ (expressed by the steps of simulation), and average force acting on the individuals during the evacuation $F_a$ – are taken into account. The model, in which the evacuation paths are related to stadium symmetry considerations, as may be seen on the figure (7) significantly minimalize the $F_a$ while the $t_e$ remain almost the same as for the traditional construction.

Acknowledgement

This work is partially supported by EU program SOCIONICAL (FP7 No 231288).